\documentclass[english,aps,prd,superscriptaddress,showpacs,preprintnumbers,amsmath,amssymb,floatfix,merge]{revtex4-1}
\usepackage[T1]{fontenc}
\usepackage[latin9]{inputenc}

\usepackage{babel}
\usepackage{graphicx}
\usepackage{multirow}
\usepackage{caption}
\usepackage{subcaption}
\usepackage{mathrsfs}
\usepackage{bbold}
\usepackage{amsmath}
\usepackage{xcolor}
\usepackage[colorlinks,linkcolor=blue,citecolor=blue,urlcolor=blue]{hyper ref}

\begin{document}

\title{Multidimensional effective field theory analysis for direct detection of dark matter}

\author{H.~E.~Rogers} \email{Corresponding author: hrogers@physics.umn.edu} \affiliation{School of Physics \& Astronomy, University of Minnesota, Minneapolis, Minnesota 55455, USA}
\author{D.~G.~Cerde\~no} \affiliation{Institute for Particle Physics Phenomenology, Department of Physics Durham University, Durham DH1 3LE, United Kingdom} \affiliation{Instituto de F\'osica Te\'orica UAM/CSIC, Universidad Aut\'onoma de Madrid, 28049, Madrid, Spain}
\author{P.~Cushman} \affiliation{School of Physics \& Astronomy, University of Minnesota, Minneapolis, Minnesota 55455, USA}	
\author{F.~Livet} \affiliation{D\'epartement Math\'ematiques, \'Ecole Normale Sup\'erieure de Cachan, 61 Avenue du Pr\'esident Wilson, 94230 Cachan, France} \affiliation{School of Physics \& Astronomy, University of Minnesota, Minneapolis, Minnesota 55455, USA}
\author{V.~Mandic} \affiliation{School of Physics \& Astronomy, University of Minnesota, Minneapolis, Minnesota 55455, USA}

\begin{abstract}
The scattering of dark matter particles off nuclei in direct detection experiments can be described in terms of a multidimensional effective field theory (EFT). A new systematic analysis technique is developed using the EFT approach and Bayesian inference methods to exploit, when possible, the energy-dependent information of the detected events, experimental efficiencies, and backgrounds. Highly dimensional likelihoods are calculated over the mass of the weakly interacting massive particle (WIMP) and multiple EFT coupling coefficients, which can then be used to set limits on these parameters and choose models (EFT operators) that best fit the direct detection data. Expanding the parameter space beyond the standard spin-independent isoscalar cross section and WIMP mass reduces tensions between previously published experiments. Combining these experiments to form a single joint likelihood leads to stronger limits than when each experiment is considered on its own. Simulations using two nonstandard operators ($\mathcal{O}_3$ and $\mathcal{O}_8$) are used to test the proposed analysis technique in up to five dimensions and demonstrate the importance of using multiple likelihood projections when determining constraints on WIMP mass and EFT coupling coefficients. In particular, this shows that an explicit momentum dependence in dark matter scattering can be identified.
\end{abstract}

\maketitle

\section{Introduction}

Astrophysical evidence indicates that nonluminous matter makes up approximately 80\% of the mass in the Universe \cite{Zwicky:1933gu,Ade:2015xua,Clowe:2003tk,Aubourg:2014yra}. If dark matter is in the form of weakly interacting massive particles (WIMPs) and interacts with baryonic matter on the scale of the weak interaction, it could be accessible to direct detection experiments sensitive to nuclear scattering from a variety of targets. The nuclear recoil energy is quite small and is expected to occur within the range of 1 to 100~keV for a WIMP mass range of 10 to 1000~GeV~\cite{Lewin:1995rx} using natural units with $c=\hbar=1$. The simplest form of the differential rate is given by
\begin{equation} \label{eq:diffRate}
{{dR} \over {dE_R}} = {{\rho_\chi} \over {m_T m_{\chi}}} \int_{v_{\text{min}}}^{\infty} v f(\vec{v}){{d \sigma} \over {d E_R}} d^3\vec{v},
\end{equation}
where $E_R$ is the nuclear recoil energy, $\rho_\chi$ is the expected local dark matter density, $m_T$ is the target nuclear mass, $m_{\chi}$ is the WIMP mass, $f(v)$ is the velocity distribution of the dark matter halo, and ${d\sigma} /{dE_R}$ is the differential cross section for the target-WIMP interaction \cite{Cerdeno:2010jj}. The minimum velocity, which is determined from nonrelativistic scattering in the center-of-mass frame, is related to the nuclear recoil energy by
\begin{equation}
v_{\text{min}} = \sqrt{m_N E_R \over 2 \mu^2},
\end{equation}
where $\mu$ is the reduced mass of the system, and $m_N$ is the mass of the neutron.

The nonrelativistic limit is considered valid for direct cold dark matter detection. The generally accepted velocity distribution for the dark matter halo is Maxwell-Boltzmann shifted by the Earth's velocity, $v_E \sim$ 232~km/s, and with the width determined by the mean velocity of the particles in the dark matter halo encompassing the galaxy, $v_0=220$~km/s. The probability of finding a dark matter particle with a velocity greater than the galactic escape velocity, $v_{\text{esc}} = 544$~km/s, is roughly zero \cite{Lewin:1995rx}. This is introduced through a cutoff to the Maxwell-Boltzmann distribution \cite{Drukier:1986tm,McCabe:2010zh}, giving the velocity distribution function of
\begin{equation}
\label{eq:vel}
f(\vec{v}) \propto e^{-(\vec{v} + \vec{v}_E)^2/v_0^2}-e^{-v_{\text{esc}}^2 / v_0^2}.
\end{equation}

The energy deposited by this nuclear recoil can be observed in the form of some combination of complementary signals of ionization, scintillation, and phonon emission depending on the target material chosen which allows for discrimination between nuclear recoil and background electron recoil events. There are many direct detection technologies and a variety of targets. Examples include noble liquids (argon, xenon, and neon), cryogenic semiconducting crystal detectors (germanium, silicon, calcium tungstate), scintillating crystal arrays (NaI, CsI), and superheated bubble chambers using a variety of fluorinated hydrocarbons. Specifics of detector sensitivity and noise determine a detector-dependent recoil energy threshold, and the discrimination afforded by complementary signals affects background discrimination \cite{Lewin:1995rx}. 

In the case of a positive signature in any of these experiments, the observed number of events and the spectral shape of the nuclear recoil spectrum can be used to determine the dark matter properties \cite{Green:2007rb}, including potential nonstandard momentum dependent contributions \cite{McDermott:2011hx}. The goodness of the reconstruction is very sensitive to uncertainties in the astrophysical parameters describing the Milky Way halo \cite{Green:2011bv} as well as in the nuclear form factors \cite{Cerdeno:2012ix} and is also subject to statistical limitations \cite{Strege:2012kv}. Regarding the WIMP-nucleus cross section, current dark matter direct detection analyses generally interpret results based on the simplest models of spin-independent or spin-dependent interactions to foster easy comparison between experiments. These conventional assumptions include form factors that are based on models of the weak force that limit the possible structure of the target nucleus and dark matter itself. It was found that, while a single experiment would be insufficient to unambiguously discriminate between spin-dependent and spin-independent couplings, a combination of targets \cite{Bertone:2007xj,Pato:2010zk,Cerdeno:2013gqa} could be used to this aim. 

Recently, a general description of the WIMP-nucleus interaction has been derived using an effective field theory (EFT) approach \cite{Fitzpatrick:2012ix,Fitzpatrick:2012ib,Anand:2013yka}. This formalism extends the model-driven conventional technique by considering all relevant couplings in the nonrelativistic limit \cite{Catena:2014uqa,DelNobile:2013sia}. The addition of angular-momentum-dependent and spin-and-angular-momentum-dependent couplings means that EFT includes interaction operators which are also dependent on momentum transfer and the initial velocities \cite{Fitzpatrick:2012ix}. The reconstruction of WIMP parameters is even more challenging in the resulting multidimensional parameter space. However, since each target nucleus is sensitive to different aspects of dark matter interactions \cite{Fitzpatrick:2012ib}, combining the results from multiple targets and techniques strongly constrains theoretical models in the absence of a detection and allows for determination of the underlying physics of the interaction once a signal is seen \cite{Peter:2013aha,Catena:2014uqa}. It has thus been argued that next generation experiments constitute an excellent tool to probe the general EFT parameter space \cite{Gluscevic:2014vga,Catena:2014epa} and identify the right theory \cite{Catena:2014hla,Gluscevic:2015sqa}. Adding information from annual modulation \cite{DelNobile:2015rmp,Witte:2016ydc} is particularly useful to identify a certain class of unconventional operators.

In this paper we further investigate the parameter reconstruction in the context of the EFT formalism. We focus on current experimental data (from the CDMS II and LUX experiments), as well as future next-generation detectors (SuperCDMS and LZ). We perform a careful treatment of the background that arises from the unique constraint of not knowing the energy dependence of the background spectra while retaining the discrimination power afforded by a binned likelihood function. We also test the reliability of confidence intervals and the Bayesian evidence for determining which parameters are relevant to dark matter experiments and to what degree we can constrain these parameters.

This article is organized as follows. In Sec.~\ref{sec:eftintro} we summarize the basic aspects of the EFT formalism. In Sec.~\ref{sec:multid-EFT} we explain the details of our numerical analysis and provide a procedure for analyzing direct dark matter detection data from one or more experiments. This procedure is then used in Sec.~\ref{sec:3d-published} to study the existing experimental data from the CDMS II and LUX experiments. Finally, in Sec.~\ref{sec:simulated} we consider a hypothetical future signature in next-generation experiments and attempt to reconstruct it in a five-dimensional subset of the EFT operators that include nontrivial dependences on momentum and spin. Our conclusions are presented in Sec.~\ref{sec:conclusions}.


\section{EFT formalism overview}\label{sec:eftintro}

\begin{table}[t]
\begin{tabular}{l | c}
\hline
\multicolumn{2}{c}{P-even, $\vec{S}_{\chi}$-independent, T-conserving} \\ \hline
$\mathcal{O}_1$ & $1$ \\
$\mathcal{O}_2$ & $(v^{\bot})^2$ \\
$\mathcal{O}_3$ & $i \vec{S}_N \cdot ({\vec{q} / m_N} \times \vec{v}^{\bot})$ \\
\hline\hline
\multicolumn{2}{c}{P-even, $\vec{S}_{\chi}$-dependent, T-conserving} \\ \hline
$\mathcal{O}_4$ & $\vec{S}_{\chi} \cdot \vec{S}_N$ \\
$\mathcal{O}_5$ & $i \vec{S}_{\chi} \cdot (\vec{q}/m_N \times \vec{v}^{\bot})$ \\
$\mathcal{O}_6$ & $(\vec{S}_{\chi} \cdot \vec{q}/m_N)(\vec{S}_N \cdot \vec{q} / m_N)$ \\
\hline\hline
\multicolumn{2}{c}{P-odd, $\vec{S}_{\chi}$-independent,T-conserving} \\ \hline
$\mathcal{O}_7$ & $\vec{S}_N \cdot \vec{v}^{\bot}$ \\
\hline\hline
\multicolumn{2}{c}{P-odd, $\vec{S}_{\chi}$-dependent, T-conserving} \\ \hline
$ \mathcal{O}_8$ & $\vec{S}_{\chi} \cdot \vec{v}^{\bot}$ \\
$ \mathcal{O}_9$ & $i \vec{S}_{\chi} \cdot (\vec{S}_N \times \vec{q} /m_N)$ \\
\hline\hline
\multicolumn{2}{c}{P-odd, $\vec{S}_{\chi}$-independent, T-violating} \\ \hline
$\mathcal{O}_{10}$ & $i \vec{S}_N \cdot \vec{q} / m_N $ \\
\hline\hline
\multicolumn{2}{c}{P-odd, $\vec{S}_{\chi}$-dependent, T-violating} \\ \hline
$\mathcal{O}_{11}$ & $i \vec{S}_{\chi} \cdot \vec{q} / m_N $ \\
\hline
\end{tabular}
\caption{EFT interaction operators of the effective interaction Lagrangian separated into categories of similar parity and WIMP spin dependence \cite{Fitzpatrick:2012ix}.}
\label{tab:Interactions}
\end{table}

All interactions considered in the dark matter EFT formalism, listed in Table \ref{tab:Interactions} by broad category, are four-fermion operators of elastic scattering between a dark matter particle ($\chi$) and a target nucleon ($N$). The effective interaction Lagrangian is expected to be of the form
\begin{equation}
\mathscr{L}_{\text{int}}=\sum_{\tau} \sum_i c_i^{\tau}\mathcal{O}_i \overline{\chi} \chi \overline{\tau} \tau,
\end{equation}
where $\tau$ can either be a sum over proton and neutron interactions or over isoscalar and isovector interactions and $i$ sums over all interaction types (operators). Here, the isoscalar/isovector basis will be used instead of the proton/neutron one. While the goal of EFT is model independence, there are some symmetries and assumptions that limit the interaction types considered, as follows. The operator variables of the effective interaction Lagrangian must have Galilean invariance. This means that the momentum- and velocity-dependent terms must appear as the momentum transfer, $\vec{q} = \vec{p}_{\chi,\text{out}} - \vec{p}_{\chi,\text{in}}$, and the relative incoming velocities, $\vec{v} = \vec{v}_{\chi,\text{in}} - \vec{v}_{N,\text{in}}$. Only elastic collisions are considered, so the kinetic energy must be conserved~\cite{Fitzpatrick:2012ib} by
\begin{equation}
{1 \over 2} \mu v^2 = {1 \over 2} \mu (\vec{v} + {\vec{q} \over \mu})^2,
\end{equation}
which leads to
\begin{equation}
\vec{v} \cdot \vec{q} = - {q^2 \over {2 \mu}}.
\end{equation}

Requiring the interaction to be Hermitian means that only four terms may appear anywhere in the effective interaction Lagrangian: the momentum transfer, $i \vec{q} / m_N$, the spin of the target, $\vec{S}_N$, the possible spin of the dark matter particle, $\vec{S}_{\chi}$, and the transverse component of the incoming velocity, $\vec{v}^{\bot} = \vec{v} + \vec{q} / 2 \mu$ \cite{Anand:2013yka}. The transverse component of the incoming velocity is chosen such that each term is linearly independent of all others. For example,
\begin{equation}
\vec{v}^{\bot} \cdot \vec{q} = 0.
\end{equation}

The EFT operators shown in Table \ref{tab:Interactions} all consist of combinations of these four terms, except for $\mathcal{O}_1$, which, as the spin-independent (SI) operator, is the simplest interaction possible. Standard SI dark matter analyses compute parameter constraints assuming that interactions with protons and neutrons are the same. This corresponds to the EFT isoscalar case defined here. EFT $\mathcal{O}_4$ is the standard spin-dependent (SD) operator and is dependent on the spin of both dark matter and the target nuclei. Past SD analyses typically assumed the proton/neutron basis instead of the isoscalar/isovector basis. Other operators, such as $\mathcal{O}_3$ and $\mathcal{O}_5$, are dependent on the momentum transfer and are characterized by different shapes of the recoil energy spectra than is typically assumed. As shown later by simulations in $\mathcal{O}_3$, experiments with low energy thresholds are particularly important for discriminating between operators associated with different spectral shapes.

If Lorentz invariance is required, then time-reversal symmetry must also be considered. Therefore, the possible interaction terms are organized by Table \ref{tab:Interactions} into T-conserving and T-violating types. The interactions are also classified by whether they have even or odd parity and if they depend on dark matter spin, $\vec{S}_{\chi}$. Within similar regions, there can be interference between operators. For example, interference terms in the Lagrangian exist between $\mathcal{O}_1$ and $\mathcal{O}_3$, $\mathcal{O}_4$ and $\mathcal{O}_5$, $\mathcal{O}_4$ and $\mathcal{O}_6$, and $\mathcal{O}_8$ and $\mathcal{O}_9$ \cite{Fitzpatrick:2012ix}.

All of the EFT operators are found as leading-order terms in the nonrelativistic reduction of a relativistic operator with a traditional spin-0 or spin-1 mediator except for $\mathcal{O}_2$. For this reason, $\mathcal{O}_2$ is not considered. Four more nonrelativistic operators exist from interactions without a spin-0 or spin-1 mediator; however, these are not linearly independent from the first eleven and are therefore not considered in order to simplify the analysis \cite{Anand:2013yka}.

Once the form factors for each interaction are known, the differential cross section is calculated as follows:
\begin{equation}
{d \sigma \over {d \cos\theta}} = {\mu^2 \over {32 \pi m_\chi^2 m_N^2}} \sum_{i,j = 1}^{11} \sum_{\tau,\tau'} c_i^{\tau} c_j^{\tau'} F_{i,j}^{\tau,\tau'}(v^2,q^2),
\end{equation}
where $c_i^{\tau}$ is the coupling coefficient for the $i^{\text{th}}$ interaction term to the nucleon or isospin. A listing of form factors for fluorine, sodium, germanium, iodine, and xenon can be found in Fitzpatrick \textit{et al} \cite{Fitzpatrick:2012ix}. This leads to a differential event rate per detector mass (cf. Eq.~\ref{eq:diffRate}) of
\begin{equation}\label{eq:diffRateEFT}
{dR \over {dE_R}} = N_T { \rho_{\chi} m_T \over {32 \pi m_{\chi}^3 m_N^2}} \int_{v_{min}}^{\infty} d^3 \vec{v} {f(\vec{v}) \over v} \sum_{i,j=1}^{11} \sum_{\tau,\tau'} c_i^{\tau} c_j^{\tau'} F_{i,j}^{\tau,\tau'}(v^2,q^2),
\end{equation}
where $N_T$ is the number of target nuclei per detector mass and the maximum WIMP velocity is encoded in the Gaussian cutoff defined in Eq.~\ref{eq:vel}.

The differential rate equation [Eq.~\ref{eq:diffRateEFT}] can be calculated as a sum over isospin (isoscalar and isovector) or as a sum over protons and neutrons. The coupling coefficients, $c_i^{\tau}$, can be converted between the nucleon and isospin bases by
\begin{equation}
c_i^0 = {1 \over 2}(c_i^p+c_i^n)
\end{equation}
and
\begin{equation}
c_i^1 = {1 \over 2}(c_i^p-c_i^n),
\end{equation}
where $c_i^0$ is the isoscalar interaction and $c_i^1$ is the isovector interaction \cite{Anand:2013yka}.

\section{Multidimensional Effective Field Theory Analysis Technique}\label{sec:multid-EFT}

In order to interpret data from direct detection experiments within the general context of EFT operators, a likelihood calculation is carried out comparing the data to theoretical models. Given the low number of expected detected WIMP events, a Poissonian likelihood function is the most reasonable choice to compare the detected recoil energy spectra with the theoretical spectra. This method has been used in maximum likelihood analyses for many different dark matter experiments. Often in dark matter analyses the likelihood function used includes only a single energy bin, sacrificing discrimination based on the recoil energy for simplicity \cite{Catena:2014uqa,DelNobile:2013sia,Catena:2014epa,Strege:2012kv,Gluscevic:2014vga,Witte:2016ydc}. In order to include the spectral information of both the expected WIMP spectrum and the detected events, the Poissonian likelihood can be split into $n$ bins, giving a dark matter-only likelihood, $\mathcal{L}^{\text{DM}}$, of
\begin{equation} \label{eq:DMonly_Likelihood}
\mathcal{L}^{\text{DM}} = \prod_{k=1}^n {1\over N_k!} \eta_k^{N_k}e^{-\eta_k},
\end{equation}
where $\eta_k$ is the expected number of events, and $N_k$ is the number of detected events in the $k^{\text{th}}$ energy bin. Binned likelihood functions have been used in some previous dark matter analyses as well \cite{Peter:2013aha,Gluscevic:2015sqa}. The expected number of events in any given energy bin is calculated using Eq.~\ref{eq:diffRateEFT} for a chosen combination of WIMP mass, $m_\chi$, and nonzero coupling coefficients; therefore,
\begin{equation} \label{eq:Event_theory}
\eta_k(\{m_\chi,c_i^0,c_i^1\}) = \int_{E_k} {dR \over {dE_R}} \; dE_R,
\end{equation}
where $i$ ranges over any operators with nonzero coupling coefficients in either isoscalar ($c_i^0$) or isovector ($c_i^1$) directions, and the integral is evaluated over the $k^{\rm th}$ energy bin $E_k$.

If the energy dependence of the backgrounds is known for an experiment, this can be added into the likelihood function as
\begin{equation} \label{eq:DMbkg_spectrum}
\mathcal{L}^{\text{DM}+\text{bkg}} = \prod_{k=1}^n {1\over N_k!} (\eta_k + b_k)^{N_k} e^{-(\eta_k+b_k)},
\end{equation}
where $b_k$ is the number of background events in the $k^{\rm th}$ energy bin. Most experiments only publish an estimate of their backgrounds across their entire energy range. Because of this, previous analyses have ignored expected background all together \cite{Peter:2013aha,Gluscevic:2014vga,Gluscevic:2015sqa,Witte:2016ydc} or only used simplistic and assumed background models \cite{Catena:2014epa,Catena:2014hla}. In order to include the background, our likelihood definition, therefore, must allow for a constraint on the total (single energy bin) background estimate, while still retaining the binned energy formalism for the WIMP data. In order to include an unbinned background in a binned likelihood, we consider all possible background configurations across energy bins that yield a total count within 2$\sigma$ of the expected total background, $B$. The background configuration that maximizes the likelihood is then accepted. The number of background configurations to be tested is determined by how many ways the total number of background counts can be distributed into all of the energy bins and is, most generally,
\begin{equation}\label{eq:combinatorics}
\text{number of background combinations} = {(n+B-1)!\over {(n-1)!B!}}.
\end{equation}
For $n = 100$ energy bins, the number of combinations necessary to test is computationally feasible for a maximum expected background of four counts or less ($B+2\sigma\le4.49$). Using this method, the likelihood can be defined for an expected total background $B$ with error $\sigma$ as
\begin{equation}\label{eq:Like_DMbkg}
\mathcal{L}^{\text{DM}+\text{bkg}} = {1\over \sqrt{2\pi\sigma}}e^{-(\sum_k^n b_k - B)^2/2\sigma^2} \prod_{k=1}^n {1\over N_k!} (\eta_k + b_k)^{N_k} e^{-(\eta_k+b_k)}.
\end{equation}

Once the likelihood has been calculated for a specific experiment, it can be combined with the likelihoods for other experiments (with potentially different targets) in order to better probe the operator space, as has been shown to be useful in previous analyses \cite{McDermott:2011hx,Strege:2012kv,Peter:2013aha}. The likelihoods are combined as
\begin{equation}\label{eq:combine_likelihoods}
\mathcal{L}_{\text{combined}} = \prod_j \mathcal{L}_{j},
\end{equation}
where $\mathcal{L}_{j}$ is the likelihood of the $j^{\rm th}$ experiment. The resulting likelihoods can be used to show the effect of each experiment (or target) on the chosen operator space and to set joint constraints on the operator space due to all available experiments. Generally, once multiplied together, the 95\% confidence contours calculated from the joint likelihood are tighter and more clearly defined.

The theoretical spectrum, $\eta_k$, is a function of the WIMP mass and all of the possible EFT coupling coefficients, thus our parameter space $\{m_\chi, c_i^0, c_i^1\}$ contains up to 23 variables. An efficient method of scanning over all possible dimensions, especially since the likelihood functions tend to be multimodal, is by using the nested sampling Monte Carlo software package \textsc{\textsc{MultiNest}} \cite{Feroz:2008xx,Feroz:2007kg,Feroz:2013hea}, a Bayesian inference tool that can be used for parameter estimation or model comparison and selection. The nested sampling technique used by \textsc{MultiNest} involves an optimized set of live points from the full likelihood. This optimized set includes the points of highest likelihood such that at each iteration of the algorithm a live point of the lowest likelihood is replaced with a point of higher likelihood \cite{Feroz:2008xx}.

Even with a software program like \textsc{MultiNest}, calculating a 23-dimensional likelihood remains computationally intensive and time consuming. Hence, exploring the likelihood over 3D subspaces corresponding to individual EFT operators (spanned by $\{m_{\chi},c_i^0,c_i^1\}$ for the operator $\mathcal{O}_i$) could be used to initially identify which operators are the most consistent with the data. For this purpose we use the Bayesian evidence, $\mathcal{Z}$, to calculate the probability that the detected data, $N_k$, is best represented by a given operator hypothesis, $H$, and is calculated by
\begin{equation}
\mathcal{Z} = \int dH\mathcal{L}(N_k | H)\text{Pr}(H),
\end{equation}
with the integral over all parameters belonging to that operator hypothesis and where $\text{Pr}(H)$ is the prior for each parameter. For the 3D example with a single operator $\mathcal{O}_i$,
\begin{equation}\label{eq:Bayesian_evidence}
\mathcal{Z}_i= \int dm_\chi dc_i^0 dc_i^1 \mathcal{L}(\{m_\chi, c_i^0, c_i^1\})\text{Pr}(\{m_{\chi},c_i^0,c_i^1\})
\end{equation}
with flat priors assumed for each parameter. The evidence is used in Bayes' theorem as
\begin{equation}\label{eq:Bayes_theorem}
\mathcal{P}(\{m_\chi,c_i^0,c_i^1\}|N_k) = {{\mathcal{L}(N_k|\{m_\chi, c_i^0,c_i^1\}) \text{Pr}(\{m_\chi,c_i^0,c_i^1\})} \over {\mathcal{Z}}},
\end{equation}
where $\mathcal{P}(\{m_\chi,c_i^0,c_i^1\}|\{N_k\})$ is the posterior probability distribution in the $\{m_\chi, c_i^0, c_i^1\}$ parameter space given the observed data, $\{N_k\}$. For a given experiment, the operators with the highest Bayesian evidence \footnote{Suppose the class of models considered consists of two models, A and B, with model A having \textit{a priori} probability $P_\text{A}$ and separate Bayesian evidence $\mathcal{Z}_\text{A}$, while model B has probability $P_\text{B} = 1-P_\text{A}$ and Bayesian evidence $\mathcal{Z}_\text{B}$. Then the data can be considered to favor model A if the posterior probability, $P'_\text{A}$, is larger than the \textit{a priori} probability, $P_\text{A}$.  Bayes' theorem gives
\begin{equation}
P'_\text{A} = \frac{P_\text{A} \mathcal{Z}_\text{A}}{P_\text{A}\mathcal{Z}_\text{A} + (1-P_\text{A})\mathcal{Z}_\text{B} }
\end{equation}
which implies $P'_\text{A} > P_\text{A}$ if and only if $\mathcal{Z}_\text{A}/\mathcal{Z}_\text{B} > 1$. If there are more than two hypotheses considered in the class of models, this holds true as long as model B includes all models except model A.} 
should be most relevant to the data and thus most likely to give nonzero coupling coefficients when analyzed jointly with other operators. The Bayesian evidence has become a fairly standard way of comparing competing models within dark matter likelihood analyses \cite{Feroz:2008xx,Catena:2014epa,Gluscevic:2014vga,Gluscevic:2015sqa,Witte:2016ydc}. In order to visualize 3D or higher dimensional likelihoods, the likelihoods can be marginalized down to multiple 2D and 1D marginalized likelihoods. Contours at 95\% confidence can be calculated in 2D planes to place constraints on likely WIMP mass and coupling coefficient values. 1D marginalized likelihoods can be used to determine the 95\% confidence regions for each parameter individually, by integrating down from the point of highest likelihood. We calculate the 95\% confidence intervals, because they have been shown to be a reliable method of estimating the true values of likelihood parameters for dark matter experiments \cite{Strege:2012kv}.

We propose the following procedure for analyzing direct dark matter detection data from one or more experiments in the vast EFT parameter space:

\begin{enumerate}
\item Assuming a flat prior for all parameters involved, run 3D analysis for each EFT operator computing the likelihood dependent only on the WIMP mass and the isoscalar and isovector coupling coefficients of that operator. \label{item:3D}
\item Calculate the Bayesian evidence, as defined in Eq.~\ref{eq:Bayesian_evidence}, for each operator's 3D likelihood. The evidence can be used to determine which operators or combination of operators are most relevant to the data set and therefore, which model best represents the dark matter interaction. \label{item:Bayes}
\item Run 5D or higher dimensional analysis for the combination of two or more relevant operators determined in step \ref{item:Bayes}, and compute constraints on the WIMP mass and relevant coupling coefficients simultaneously over all relevant EFT operators. \label{item:5D}
\item Combine the likelihoods of individual experiments for relevant operators into one likelihood. The joint likelihood can be used to compute the most stringent constraints on EFT parameters, using information from all available experiments. \label{item:Targets}
\end{enumerate}

\section{3D Analysis of Published Data}\label{sec:3d-published}
\begin{table}[b]
\begin{tabular}{c|c|c|c}
Experiment & Exposure (kg days) & Events (keV) & Background (counts) \\ \hline
CDMS II Si~\cite{Agnese:2013rvf} & 140.2 & 8.2, 9.5, 12.3 & 0.41$\pm$0.48 \\
CDMS II Ge~\cite{Agnese:2015ywx} & 612 & 10.81, 12.3 & 0.64$\pm$0.17 \\
LUX~\cite{Akerib:2013tjd} & 10065.4 & $\sim$4.5 & 0.64$\pm$0.16 \\
\end{tabular}
\caption{Overview of the published results from each of the chosen experiments.}
\label{tab:published_overview}
\end{table}

\begin{figure}[t]
\centering
\includegraphics[width = \columnwidth]{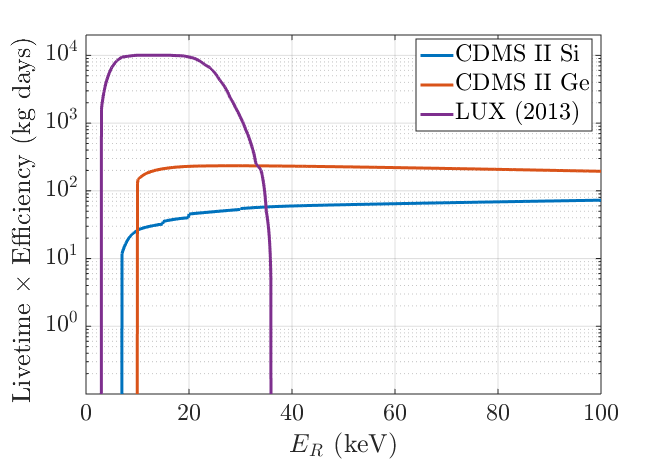}
\caption{Exposures of CDMS II Si~\cite{Agnese:2013rvf}, CDMS II Ge~\cite{Agnese:2015ywx}, and LUX~\cite{Akerib:2013tjd} as a function of recoil energy including the experimental efficiencies and energy thresholds.}
\label{fig:real_data_exposure}
\end{figure}

The EFT analysis methodology described above can be used to present new interpretations of previously published WIMP search results. To demonstrate this, consider three past results obtained using experiments with different target materials: Cryogenic Dark Matter Search (CDMS) experiment observed three WIMP candidate events using silicon detectors \cite{Agnese:2013rvf} and two using germanium detectors~\cite{Agnese:2015ywx}, while the Large Underground Xenon (LUX) experiment observed one candidate WIMP event using a liquid xenon detector~\cite{Akerib:2013tjd}. While not the most recent results in the field, they were chosen to illustrate an example of a tension between different experiments. All three measurements assumed an isoscalar spin-independent interaction (cross section, $\sigma_1^0$) and published results for a range of WIMP masses ($m_\chi$) based on detected nuclear recoil events. The energies of the detected events, total exposure, and expected background for each experiment are shown in Table \ref{tab:published_overview}. A comparison of the exposures of each experiment, including the efficiencies and energy thresholds, is shown in Fig.~\ref{fig:real_data_exposure}.

\begin{figure}[t]
\centering
\includegraphics[width = \columnwidth]{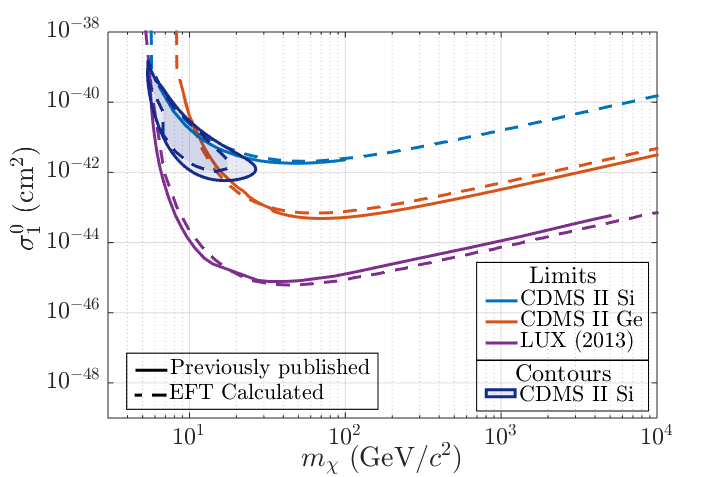}
\caption{EFT 95\% upper limit contours for each experiment and the silicon 95\% detection contour, which were calculated in a 2D likelihood analysis of WIMP mass and the isoscalar operator 1 cross section, are compared to the published optimal interval contours for the CDMS experiments~\cite{Agnese:2013rvf,Agnese:2015ywx} and to the profile likelihood ratio upper limit from LUX~\cite{Akerib:2013tjd}. The LUX limit (purple) rules out the CDMS II Si contour (blue).}
\label{fig:EFT_c10_target_compare}
\end{figure}

Using these parameters, the results obtained by the three experiments were reproduced using the EFT SI operator. Figure \ref{fig:EFT_c10_target_compare} compares the previously published results to the constraints calculated using the EFT likelihood analysis technique for each experiment over the two-dimensional parameters space of the WIMP mass and the elastic scattering cross section due to the isoscalar component of operator 1, $\sigma_1^0$. The published upper limits from all three experiments and the detection contour for the CDMS II Si result are in good agreement with the corresponding constraints obtained with the EFT likelihood analysis. In both cases, the LUX limit, shown in purple, completely excludes the CDMS II Si contour, shown in blue, leading to a visible tension between these experiments in the low mass region. Previous dark matter simulations \cite{McDermott:2011hx,Catena:2014hla} have shown that assuming the incorrect model for dark matter-target interactions can lead to biased contours and can cause tension between experiments.

\begin{figure}[t]
\centering
\begin{subfigure}{\columnwidth}
\includegraphics[width = \columnwidth]{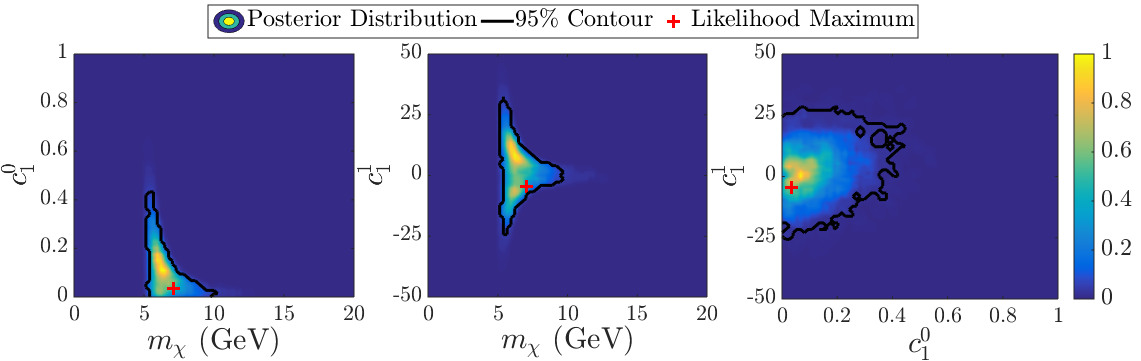}
\caption{CDMS II Si~\cite{Agnese:2013rvf}}
\label{fig:EFT_3D_Op1_real_Si}
\end{subfigure}
\begin{subfigure}{\columnwidth}
\includegraphics[width = \columnwidth]{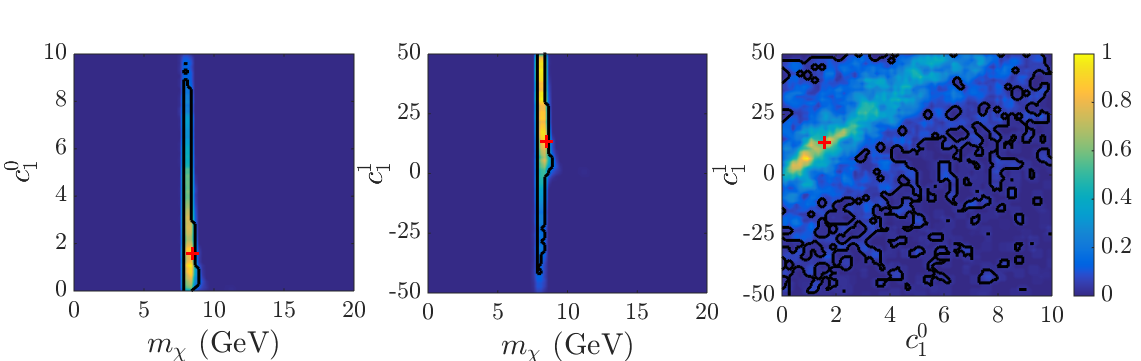}
\caption{CDMS II Ge~\cite{Agnese:2015ywx}}
\label{fig:EFT_3D_Op1_real_Ge}
\end{subfigure}
\begin{subfigure}{\columnwidth}
\includegraphics[width = \columnwidth]{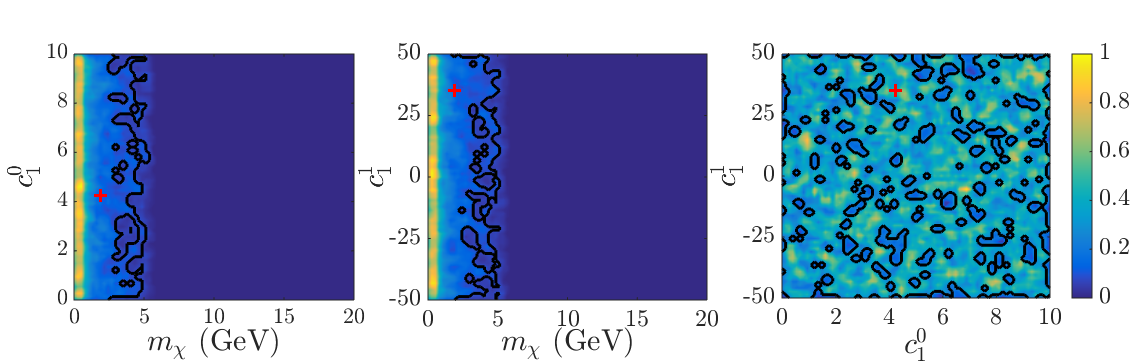}
\caption{LUX (2013)~\cite{Akerib:2013tjd}}
\label{fig:EFT_3D_Op1_real_Xe}
\end{subfigure}
\caption{2D marginalized likelihoods from the 3D likelihood of each previously published experiment calculated using WIMP mass ($m_\chi$) and both isoscalar ($c_1^0$) and isovector ($c_1^1$) coupling coefficient components of operator 1. Contours are calculated at the 95\% confidence level, and the global likelihood maximum is depicted.}
\label{fig:EFT_3D_Op1_real}
\end{figure}

\begin{figure}[t]
\centering
\includegraphics[width = \columnwidth]{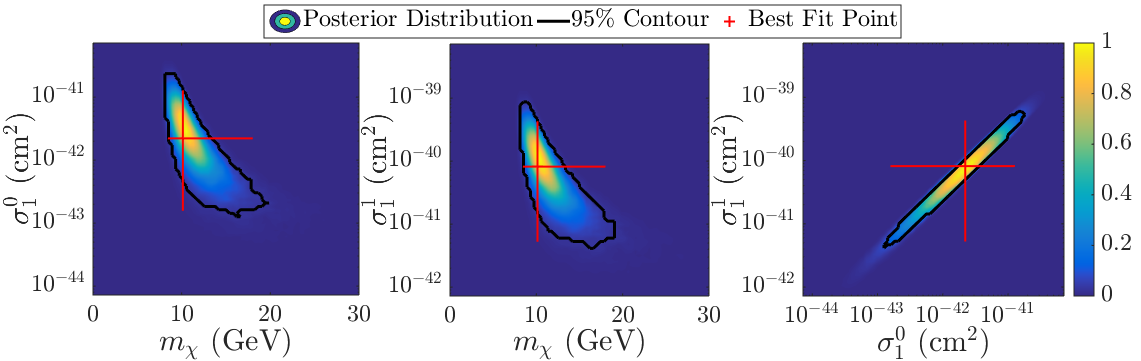}
\includegraphics[width = \columnwidth]{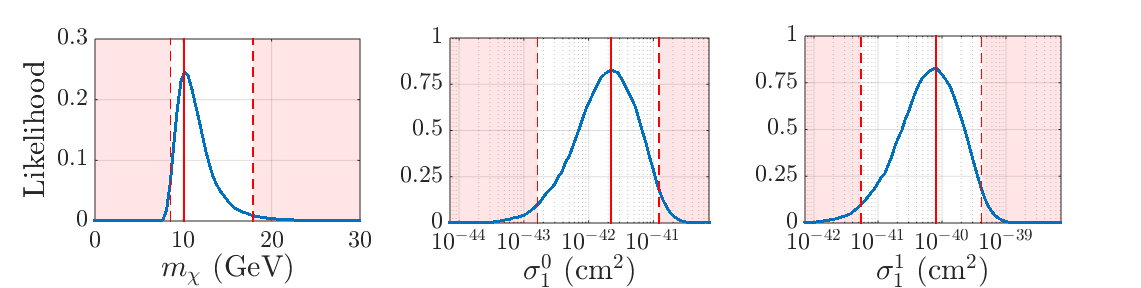}
\caption{Joint 3D likelihood combining CDMS II Si~\cite{Agnese:2013rvf}, CDMS II Ge~\cite{Agnese:2015ywx}, and LUX~\cite{Akerib:2013tjd} data. Plotted are WIMP mass ($m_\chi$), isoscalar operator 1 cross section ($\sigma_1^0$), and isovector operator 1 cross section ($\sigma_1^1$). The top row depicts 2D marginalized likelihoods obtained by marginalizing over one of the parameters, while the bottom row shows 1D marginalized likelihoods obtained by marginalizing over two of the three parameters. Also shown are the 95\% confidence contours and the point of best fit with error bars derived from the 1D marginalized likelihoods.}
\label{fig:3D_Op1_real_combined}
\end{figure}

The tension between these experiments can be relieved by generalizing the WIMP-nucleus interaction, thus including more EFT operators in the analysis. The simplest addition is the inclusion of the isovector component of $\mathcal{O}_1$ to the SI interaction leading to a likelihood calculated over the 3D parameter space $\{m_\chi,c_1^0,c_1^1\}$. We then marginalize over one of the parameters to compute 2D marginalized likelihoods for each experiment individually, as shown in Fig.~\ref{fig:EFT_3D_Op1_real}. The 95\% confidence contours shown for CDMS II Ge and LUX are open contours, consistent with the published LUX and CDMS limits. The symmetries visible in the likelihoods, especially in CDMS II Ge, indicate that the isoscalar and isovector components have the same sign. Combining all three experiments together into a single likelihood makes this symmetry stronger, such that no negative values of the isovector coupling coefficient remain. Figure \ref{fig:3D_Op1_real_combined} shows the joint likelihood (with all three experiments combined) marginalized over one of the parameters (top row) or over two of the three parameters (bottom row). The cross section is plotted instead of the coupling coefficient, defined by
\begin{equation}\label{eq:c_to_sigma}
\sigma_1^\tau = \frac {(A m_N)^2}{4 \pi \langle V\rangle^4 (1+A)^2} (c_1^\tau)^2,
\end{equation}
where $A$ is the number of nucleons of the target material and $\langle V\rangle=246.2$~GeV is the Higgs vacuum expectation value, used here to represent the electroweak scale and to define dimensionless coefficients~\cite{Anand:2013yka}.

\begin{figure}[t]
\centering
\includegraphics[width=\columnwidth]{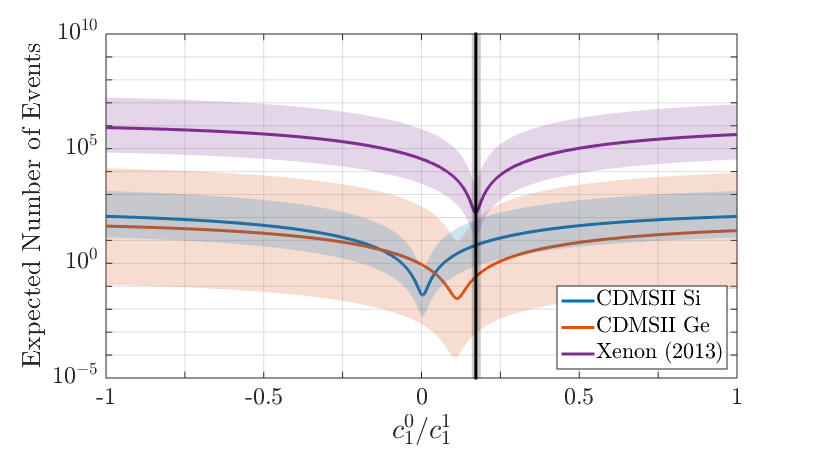}
\caption{Total integrated rate for each experiment over a range of coupling coefficient ratios for $\mathcal{O}_1$ calculated at the best fit mass of $m_\chi = 10.1$~GeV and total coupling coefficient amplitude of $\sqrt{(c_1^0)^2+(c_1^1)^2} = 0.12\pm^{0.20}_{0.09}$. The minimum for each experiment represents the ratio for which that experiment would detect the fewest number of events. The best fit point of the combined 3D likelihood is shown with 95\% confidence errors by the vertical line close to the minimum for LUX. The width of the rate for each experiment was calculated from the 95\% confidence regions of the best fit mass and of the total coupling coefficient amplitude.}
\label{fig:real_data_op1_scan}
\end{figure}

The best fit point of the joint likelihood is shown in Fig.~\ref{fig:3D_Op1_real_combined} in red with 95\% confidence intervals as calculated from the 1D marginalized likelihoods. The parameters of this point with 95\% confidence intervals are $m_\chi = 10.1 \pm^{7.8}_{1.5}$~GeV, $\sigma_1^0 = (2.2\pm^{10.1}_{2.1})\times10^{-42}\text{ cm}^2$, and $\sigma_1^1 = (8.2\pm^{33.2}_{7.6})\times10^{-41}\text{ cm}^2$. The ratio between coupling coefficients of the best fit point, $c_1^0/c_1^1 = 0.172\pm^{0.016}_{0.013}$, coincides with the point for which the sensitivity of LUX is at the lowest, as shown in Fig.~\ref{fig:real_data_op1_scan}, showing that the LUX result constrains the combined likelihood the most. The 95\% or 2$\sigma$ confidence contours around the best fit point are closed, as shown in Fig.~\ref{fig:3D_Op1_real_combined}; however, at 5$\sigma$ confidence, the contours are open, so we make no claim of dark matter detection. 
 
The Bayesian evidence can be used to evaluate whether the goodness of fit was improved by adding an isovector component. The evidence for each experiment in both the 2D and 3D analyses are shown in Table \ref{tab:real_evidence}. For each experiment individually the evidence is greater when both coupling components are included, indicating that the 3D model is a better fit than the simple 2D WIMP mass and isoscalar coupling model.

\begin{table}[t]
\begin{tabular}{c|c|c|c}
Model & CDMS II Si~\cite{Agnese:2013rvf} & CDMS II Ge~\cite{Agnese:2015ywx} & LUX~\cite{Akerib:2013tjd} \\ \hline
$c_1^0$ only & $3.54\times10^{-6}$ & $1.99\times10^{-4}$ & 0.00365 \\
$c_1^0$ and $c_1^1$ & $2.84\times10^{-5}$ & $4.38\times10^{-4}$ & 0.0104 \\
3D / 2D & 8.02 & 2.20 & 2.84 \\
\end{tabular}
\caption{Bayesian evidence for each experiment and for the two models: isoscalar spin-independent coupling only (the typically assumed case, 2D) and isoscalar and isovector spin-independent coupling (3D). For all three experiments, the evidence favors coupling via a combination of both isoscalar and isovector couplings as shown by the ratio between the 3D and 2D cases.}
\label{tab:real_evidence}
\end{table}

\section{Simulated Dark Matter Data from Future Experiments}\label{sec:simulated}

Assuming a wrong operator for WIMP-nucleon coupling when conducting an analysis of WIMP search data can lead to erroneous conclusions about the WIMP mass and interactions. Possible failure modes are demonstrated with a set of simulated experiments where the WIMP-nucleon interaction proceeds via nonstandard operators. Three hypothetical direct dark matter experiments are defined. The silicon (Si) and germanium (Ge) experiments are based on the proposed SuperCDMS SNOLAB~\cite{Agnese:2016cpb} experiment with backgrounds given by Poissonian errors of 1~count/year for 400~kg of Ge and 0.86~counts/year for 170~kg of Si. The liquid xenon (LXe) experiment is based on the LUX upgrade with a low threshold~\cite{Akerib:2015rjg} and background~\cite{Akerib:2014rda} over a live time of 33500~kg~days \cite{Akerib:2016vxi}. An overview of the assumed backgrounds, exposures, and energy thresholds is given in Table \ref{tab:sim_exp_overview}. The efficiencies are assumed to be a simple step function between the experimental threshold and the energy at which the experiment's efficiency drops back to 75\%.

\begin{table}[t]
\begin{tabular}{c|c|c}
Target & Live time (kg days) & Total background (counts) \\ \hline
Si & 63000 & 0.86$\pm$0.93 \\
Ge & 145000 & 1$\pm$1 \\
LXe & 33500 & 3.5$\pm$0.4 \\
\end{tabular}
\caption{Details used to build the simulated data for each target chosen. Each simulated experiment is assigned an energy threshold of 1~keV. The Si and Ge experiments are based on the proposed SuperCDMS SNOLAB~\cite{Agnese:2016cpb}, and the LXe on the most recent results from LUX \cite{Akerib:2016vxi,Akerib:2015rjg,Akerib:2014rda}.}
\label{tab:sim_exp_overview}
\end{table}

\begin{table}[t]
\begin{tabular}{c|c|c|c|c}
Benchmark point & $m_\chi$ (GeV) & $\{c_1^0,c_1^1\}$ & $\{c_3^0,c_3^1\}$ & $\{c_8^0,c_8^1\}$ \\ \hline
BP$_8$ & 3.0 & \{0,0\} & \{0,0\} & \{4.875,24.375\} \\
BP$_3$ & 8.0 & \{0,0\} & \{16, -6.4\} & \{0,0\} \\
\end{tabular}
\caption{WIMP mass and coupling coefficients for $\mathcal{O}_1$, $\mathcal{O}_3$, and $\mathcal{O}_8$ as benchmark points to simulate the detected dark matter data.}
\label{tab:sim_parameters}
\end{table}

We present two simulations, one in which the WIMP-nucleon scattering proceeds via EFT $\mathcal{O}_8$ and the other in which the WIMP-nucleon scattering proceeds via $\mathcal{O}_3$. In each case, the values for the WIMP mass and isoscalar and isovector coupling coefficients are chosen in order to compute the theoretical recoil energy spectra for each of the three simulated experiments. The parameters chosen for each benchmark experiment are listed in Table \ref{tab:sim_parameters}. Treating the recoil energy spectra as probability density functions, we randomly draw WIMP-event recoil energies, with the number of events in each simulated experiment given by the integral of the theoretical recoil energy spectrum. The energies of the simulated background events were randomly drawn from a flat probability density function over the energy range set by the efficiency. The simulated dark matter events and simulated background events together were used as the detected events for each simulated experiment.

\subsection{5D Analysis of Data Simulated in Operator 8}

The EFT operator $\mathcal{O}_8$, described by $\vec{S}_\chi \cdotp \vec{v}^{\perp}$, is dependent on the WIMP spin (here assumed to be $S_\chi$=1/2), the transverse component of relative velocity ($\vec{v}^\perp$), and spin-independent and angular-momentum-dependent target nuclear responses. The spin-independent nuclear response is the same as that found in the standard SI interaction, $\mathcal{O}_1$~\cite{Fitzpatrick:2012ix}. Therefore, $\mathcal{O}_8$ and $\mathcal{O}_1$ have the same exponential recoil energy spectral shape; however, the overall rate depends on the WIMP mass and does so differently depending on the target material. This operator was chosen in order to illustrate the challenge of identifying the correct WIMP-nucleon interaction operator when the operator yields similar recoil energy spectral shape to $\mathcal{O}_1$, and the only target-dependent modifier is the overall integrated rate.

We consider the benchmark point, BP$_8$, with parameters as defined in Table \ref{tab:sim_parameters}. This example was chosen specifically to produce a distinctive signal in Si but not in Ge or LXe.  The chosen ratio of isoscalar to isovector components, $c^0_8/c^1_8 = 0.2$, favors interactions with Si over Ge. Also, the low WIMP mass of 3~GeV is below the experimental threshold assumed for LXe. For exposures considered in Table \ref{tab:sim_exp_overview}, this resulted in 11, 1, and 0 events for Si, Ge, and LXe, respectively. This corresponds to 12, 2, and 4 events when the background is included.

\begin{figure}[t]
\centering
\begin{subfigure}{0.325\columnwidth}
\includegraphics[width = \columnwidth]{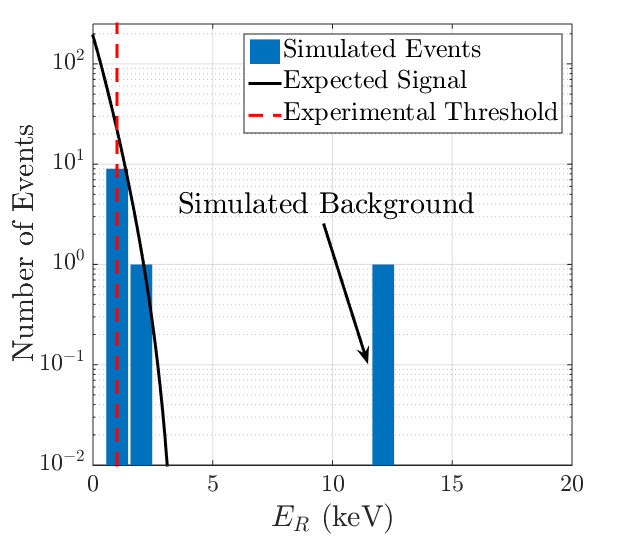}
\caption{Si}
\label{fig:op8_simulated_data_Si}
\end{subfigure}
\begin{subfigure}{0.325\columnwidth}
\includegraphics[width = \columnwidth]{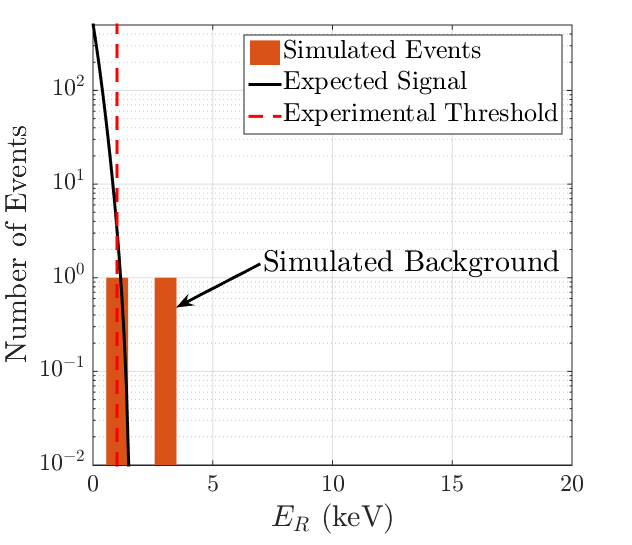}
\caption{Ge}
\label{fig:op8_simulated_data_Ge}
\end{subfigure}
\begin{subfigure}{0.325\columnwidth}
\includegraphics[width = \columnwidth]{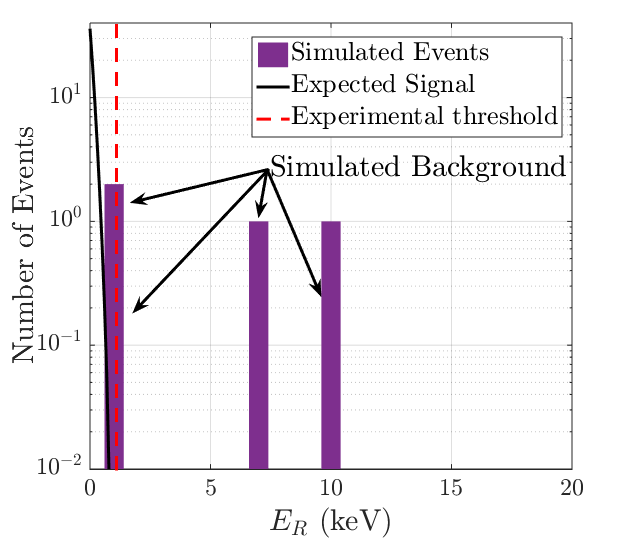}
\caption{LXe}
\label{fig:op8_simulated_data_Xe}
\end{subfigure}
\caption{The total (signal plus background) recoil energy spectra simulated for each experiment compared to the expected rates of WIMP-nucleon scattering for the chosen interaction parameters in the $\mathcal{O}_8$ simulation (BP$_8$). The dashed line indicates the energy threshold used in the simulation.}
\label{fig:op8_simulated_data}
\end{figure}

The simulated WIMP events for the three experiments are shown in Fig.~\ref{fig:op8_simulated_data}. All of the simulated events for Ge sit right at the experimental threshold, so very little shape information is available. On the contrary, for Si some simulated WIMP events pass the experimental threshold, so the shape information should be more helpful in distinguishing between operators. Even though all of the simulated data for LXe are background events, they mimic the energy distribution of an exponentially decaying WIMP spectrum, which allows the background events to be easily misinterpreted as a WIMP signal.

\begin{figure}[t]
\centering
\begin{subfigure}{\columnwidth}
\includegraphics[width = 0.875\columnwidth]{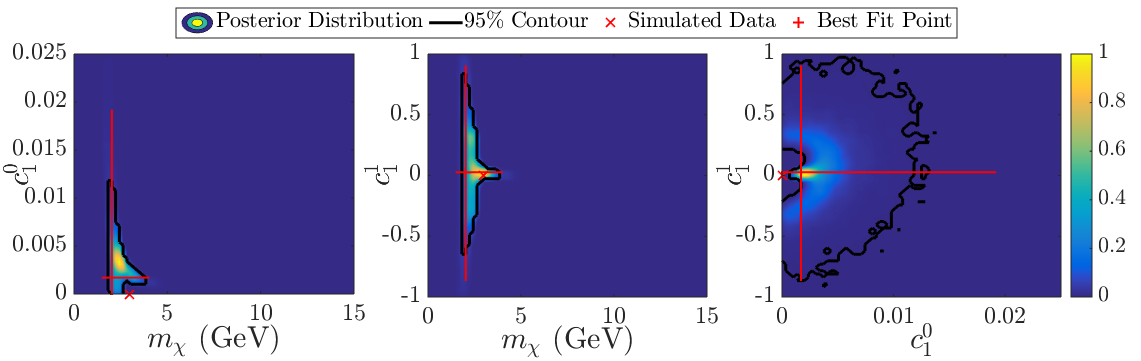}
\includegraphics[width = 0.875\columnwidth]{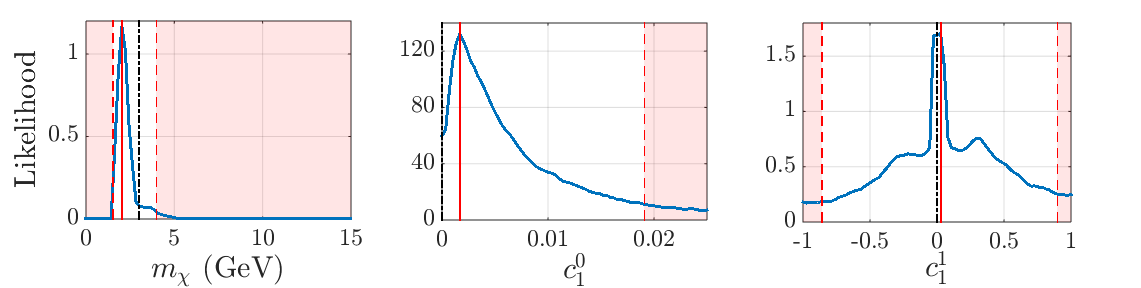}
\caption{$\mathcal{O}_1$ recovery}
\label{fig:op8_3Dlikelihoods_op1}
\end{subfigure}
\begin{subfigure}{\columnwidth}
\includegraphics[width = 0.875\columnwidth]{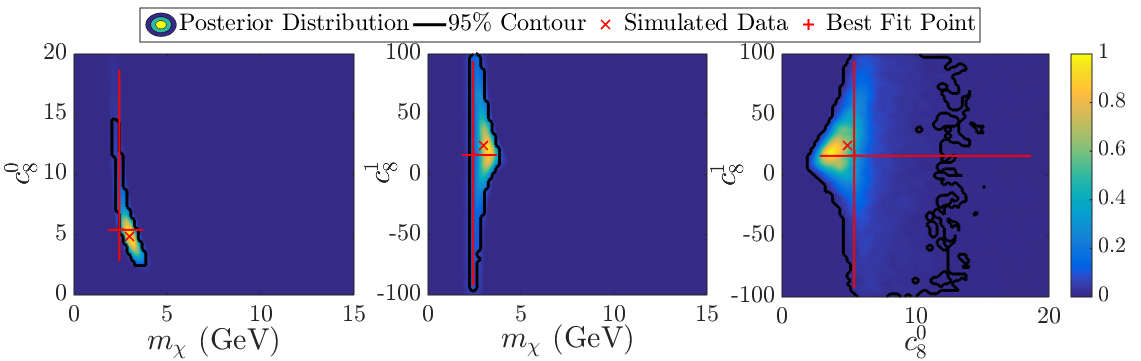}
\includegraphics[width = 0.875\columnwidth]{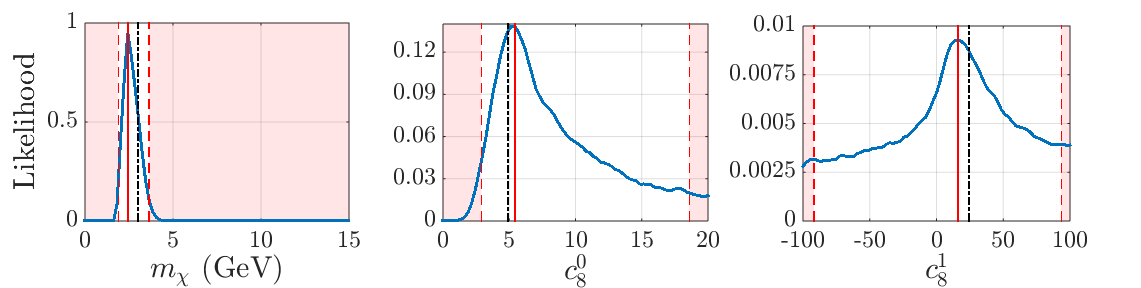}
\caption{$\mathcal{O}_8$ recovery}
\label{fig:op8_3Dlikelihoods_op8}
\end{subfigure}
\caption{3D likelihoods of the data simulated using BP$_8$ and analyzed under the assumption of $\mathcal{O}_1$ (\ref{fig:op8_3Dlikelihoods_op1}) or $\mathcal{O}_8$ (\ref{fig:op8_3Dlikelihoods_op8}) for all three experiments combined. For each recovery operator, the top row of plots shows 2D marginalized likelihoods (obtained by marginalizing over one of the parameters) and the bottom row shows the 1D marginalized likelihoods (obtained by marginalizing over two of the three parameters). Also shown is the point representing the simulated data, marked by x in 2D and a black dashed line in 1D and the best fit point represented by the red + in 2D and red vertical line in 1D.}
\label{fig:op8_3Dlikelihoods}
\end{figure}

We start by analyzing the simulated data in the EFT likelihood formalism assuming $\mathcal{O}_1$ interaction only; that is, the likelihood is computed over the 3D parameter space of $\{m_{\chi},c_1^0,c_1^1\}$. The resulting 3D likelihood is shown in Fig.~\ref{fig:op8_3Dlikelihoods_op1} with the 2D marginalized likelihoods shown on top and the 1D marginalized likelihoods on the bottom. The best-fit point, which is calculated from the 1D marginalized likelihoods and is listed in Table \ref{tab:best_fit_points_8}, is also depicted along with the error bars. The point representing the simulated data ($c_1^0=0,c_1^1=0$) is contained within the 1D 95\% confidence intervals but not in two of the 2D 95\% confidence contours. For example, the $c_1^0$ vs.~$c_1^1$ contour plot on the far top right of Fig.~\ref{fig:op8_3Dlikelihoods_op1} shows that the simulated point is not contained within the 95\% confidence contour. This example demonstrates the fact that marginalizations to one dimension, with their necessary loss of information, can be misleading. The 2D representation must be used in order to develop a better understanding of the parameter space. In this particular simulation the 2D marginalized likelihoods indicate that nonzero $\mathcal{O}_1$ couplings are needed in order to explain the simulated data. This, of course, is not consistent with the assumed simulation parameters, and it is a consequence of the fact that a wrong operator was used to analyze the data. In other words, assuming the wrong operator when calculating the likelihood can lead to reasonable 2D contours that do not represent the true (in this case, simulated) nature of dark matter.

\begin{table}[t]
\begin{tabular}{c|c|c|c}
Reconstructed point & $m_\chi$ (GeV) & $\{c_1^0,c_1^1\}\times10^3$ & $\{c_8^0,c_8^1\}$ \\ \hline \hline
\multirow{2}{*}{BP$_8$ in $\mathcal{O}_1$} & $2.0$ & $\{1,30\}$ & \multirow{2}{*}{...} \\
& $(1.5,4.0)$ & $\{(0,20),(-860,900)\}$ &  \\ \hline
\multirow{2}{*}{BP$_8$ in $\mathcal{O}_8$} & $2.4$ & \multirow{2}{*}{...} & $\{5,16\}$ \\
& $(1.9,3.7)$ & & $\{(3,19),(-92,94)\}$ \\ \hline
\multirow{2}{*}{BP$_8$ in $\mathcal{O}_1$ and $\mathcal{O}_8$} & $2.03$ & $\{2,30\}$ & $\{0,\text{--}\}$ \\
& $(1.56,2.85)$ & $\{(0,18),(-820,870)\}$ &  $\{(-18,18),(-95,95)\}$ \\
\end{tabular}
\caption{Best fit points with 95\% confidence regions for the 3D and 5D reconstructions of the benchmark point BP$_8$ of Table \ref{tab:sim_parameters}, based on 1D marginalized likelihoods. As noted in the header, $\mathcal{O}_1$ coupling coefficients have been enlarged by $10^3$.}
\label{tab:best_fit_points_8}
\end{table}

This analysis is then repeated assuming $\mathcal{O}_8$ interaction only, and the likelihood is computed over the 3D parameter space of $\{m_{\chi},c_8^0,c_8^1\}$, as shown in Fig.~\ref{fig:op8_3Dlikelihoods_op8}. In this case, the simulated point is well within the 95\% confidence 2D contours and 1D intervals, as one would expect since this recovery assumes the correct operator. The resulting likelihood is well defined in WIMP mass and isoscalar coupling coefficient but less so in the isovector component. The 95\% confidence intervals computed from 1D marginalized likelihoods are also shown in Table \ref{tab:best_fit_points_8}. These intervals were calculated using the joint (Si, Ge, and LXe) likelihood and are tighter than for any single experiment alone. Specifically, since Si detected the largest number of events (11 events versus 1 for Ge and 0 for Xe), the Si-only likelihood is expected to best match the results of the joint likelihood. However, the widths of the 95\% confidence contours were $\sim1.4$ times larger for Si alone than for the joint likelihood case, demonstrating that combining experiments tightens the resulting contours.

\begin{figure}[t]
\centering
\includegraphics[width = \columnwidth]{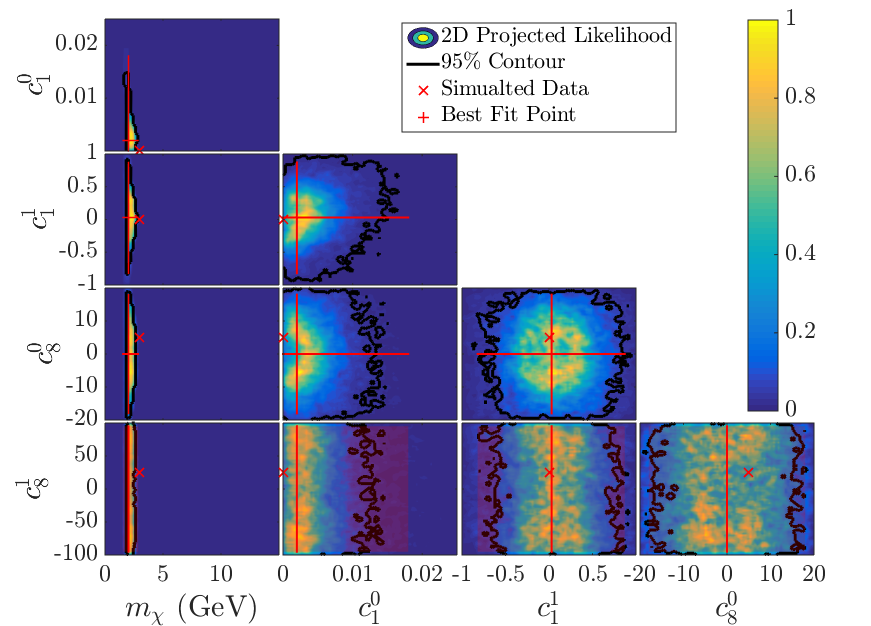}
\includegraphics[width = \columnwidth]{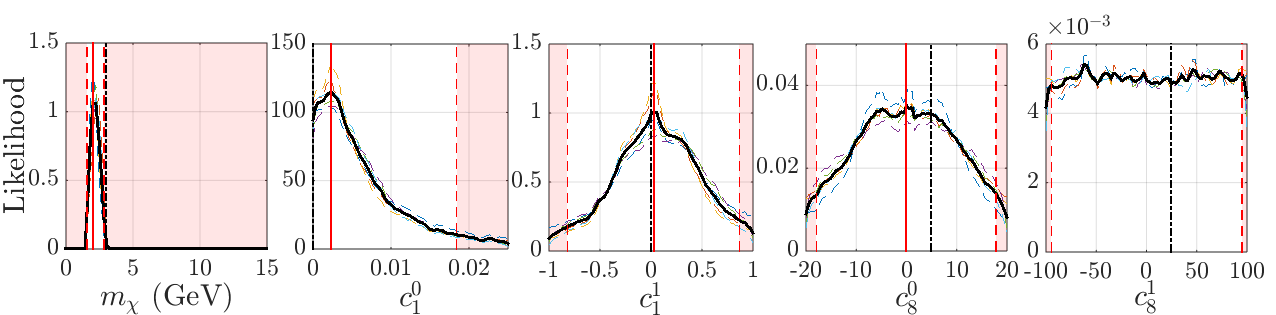}
\caption{5D likelihood of the data simulated in $\mathcal{O}_8$ and recovered assuming interactions in both $\mathcal{O}_1$ and $\mathcal{O}_8$, for all three experiments combined. The 95\% confidence contours in 2D marginalized likelihoods are shown on the top and the 1D marginalized likelihoods are shown in the bottom row of plots. The multiple colors in the 1D plot represent the marginalizations of the six subspaces and the black line the averaged. Also shown is the point representing the simulated data, marked by x in 2D and a black dashed line in 1D and the best fit point represented by the red + (or shaded red region) in 2D and red vertical line in 1D.} 
\label{fig:op8_5Dlikelihoods}
\end{figure}

Additional information can be gleaned from the Bayesian evidence. From the 3D likelihoods, the evidence for $\mathcal{O}_1$ is $2\times10^{-9}$, whereas the evidence for $\mathcal{O}_8$ is 3 times larger at $6\times10^{-9}$. This shows that $\mathcal{O}_8$ is the better fit to this data. 

In the proposed procedure for analysis of WIMP search data, step \ref{item:5D} proposes a likelihood analysis in higher-dimensional parameter space including operators with the highest evidences in 3D likelihood analyses. Applying this approach to our simulation, we perform the EFT likelihood analysis of the 5D parameter space $\{m_{\chi},c_1^0,c_1^1,c_8^0,c_8^1\}$. The results are shown in Fig.~\ref{fig:op8_5Dlikelihoods} with the 2D marginalized likelihoods on the top and the 1D marginalized likelihoods on the bottom. The 1D marginalizations of the 5D likelihood were calculated from six different 3D subspaces and averaged together to give one 1D likelihood. The six marginalizations from different 3D subspaces are plotted in varying colors in Fig.~\ref{fig:op8_5Dlikelihoods} (bottom) with the averaged curve in black. Note that all marginalized likelihoods (for a given parameter) are similar, indicating that the possible systematic error in this marginalization procedure is not significant.

The parameters of the best fit point calculated from the 1D likelihoods are found in Table \ref{tab:best_fit_points_8}. Note that the open contour for $c_8^1$ implies a flat spectrum with no discernible peak. The simulated data point is contained within all of the 95\% confidence intervals except for WIMP mass. The WIMP mass sits just outside the 95\% confidence (or 2$\sigma$ confidence) contour at 2.4$\sigma$ or at 1.6\% probability of occurring. One factor that could contribute to this are the two LXe background events just above threshold that mimic a low mass WIMP. 

Since the spectral shapes for $\mathcal{O}_1$ and $\mathcal{O}_8$ are both exponentially decaying, it is difficult to separate the four coupling coefficients from each other using only three target materials. This is most apparent in the 1D projections of the likelihood, where the peaks of $c_1^1$ and $c_8^0$ are very wide, and the likelihood for $c_8^1$ is completely flat. In other words, although the 5D likelihood analysis detects the WIMP and places a strong constraint on the WIMP mass (consistent with the simulated WIMP mass), it cannot constrain the individual couplings in $\mathcal{O}_1$ and $\mathcal{O}_8$ due to their degeneracies. Additional experiments with different targets would be needed to break these degeneracies.

\subsection{5D Analysis of Data Simulated in Operator 3}

\begin{figure}[t]
\centering
\begin{subfigure}{0.325\columnwidth}
\includegraphics[width = \columnwidth]{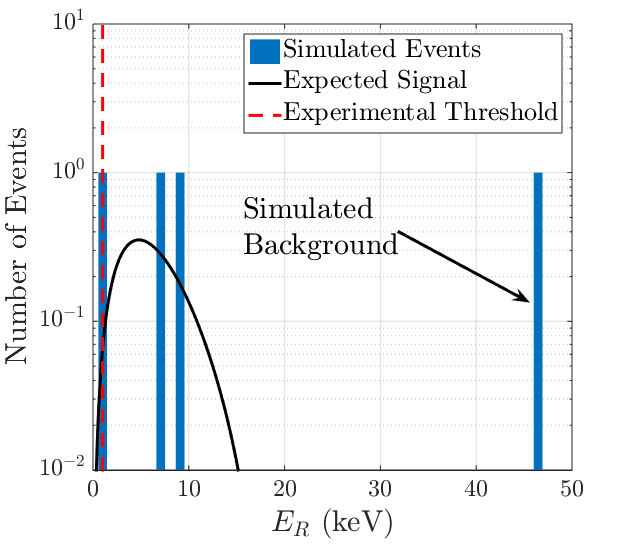}
\caption{Si}
\label{fig:op3_simulated_data_Si}
\end{subfigure}
\begin{subfigure}{0.325\columnwidth}
\includegraphics[width = \columnwidth]{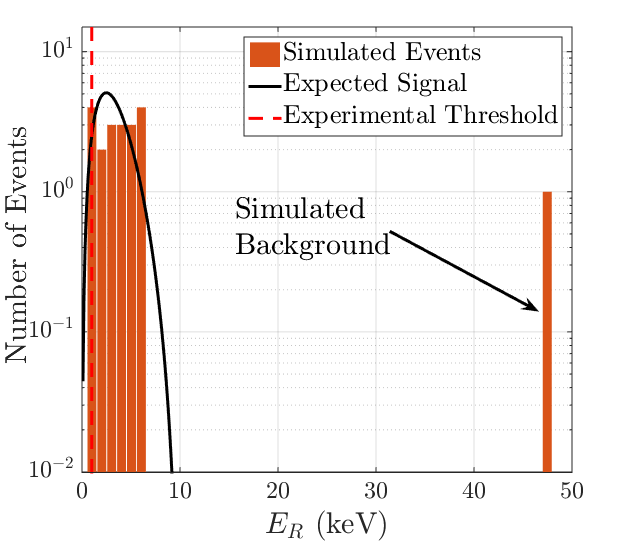}
\caption{Ge}
\label{fig:op3_simulated_data_Ge}
\end{subfigure}
\begin{subfigure}{0.325\columnwidth}
\includegraphics[width = \columnwidth]{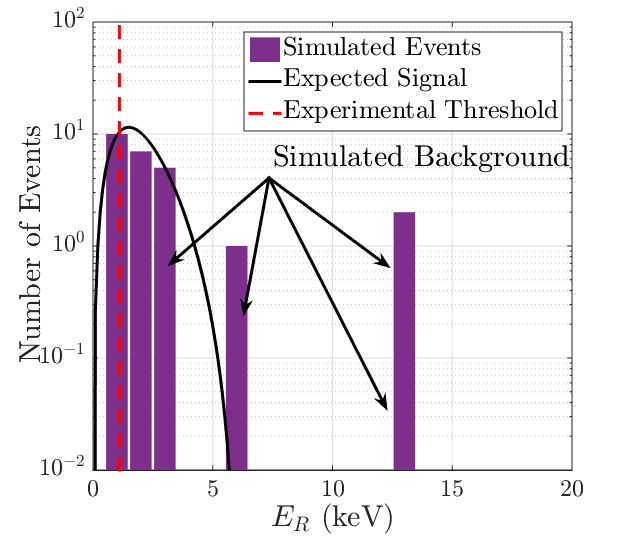}
\caption{LXe}
\label{fig:op3_simulated_data_Xe}
\end{subfigure}
\caption{The total (signal plus background) simulated data for each experiment compared to the expected recoil energy spectra of WIMP-nucleon scattering for the chosen $\mathcal{O}_3$ interaction parameters using BP$_3$. The dashed line indicates the energy threshold used in the simulation.}
\label{fig:op3_simulated_data}
\end{figure}

The EFT operator $\mathcal{O}_3$ is given by $i\vec{S}_N \cdot(\vec{q}\times\vec{v}^\perp$), has no dependence on the WIMP spin, and relies on two nuclear responses of the target: a spin-dependent response (transverse to the momentum transfer) and a spin-and-angular-momentum-dependent response~\cite{Fitzpatrick:2012ix}. Therefore, the event rate spectrum of $\mathcal{O}_3$ has a different shape than that of $\mathcal{O}_1$. In particular, the event rate spectrum for $\mathcal{O}_1$ smoothly decays exponentially with recoil energy, while for $\mathcal{O}_3$, even with no experimental efficiencies included, the event rate is suppressed at low energies with a pronounced peak at higher energies, as shown in Fig.~\ref{fig:op3_simulated_data}. The energy and amplitude of the peak is dependent on the WIMP mass, the combination of coupling coefficients, and the target chosen. This operator was chosen to demonstrate how differences in the shape of recoil energy spectra can be used to improve parameter estimation.

The three benchmark experiments (Ge, Si, LXe) are simulated in the $\mathcal{O}_3$ framework using the benchmark point BP$_3$ listed in Table \ref{tab:sim_parameters} with the ratio of isoscalar and isovector components of $c_3^0/c_3^1 = -2.5$. For the simulated exposures and energy ranges described in Table \ref{tab:sim_exp_overview}, Si detected three WIMP events, Ge detected 19 events, and LXe detected 21 events. Including the simulated background, the total number of simulated detected events for each experiment is 4, 20, and 25 events respectively. The simulated data compared to the expected recoil energy spectra for each experiment are shown in Fig.~\ref{fig:op3_simulated_data}.

The numbers of simulated events for Ge and LXe (Fig.~\ref{fig:op3_simulated_data}) are large enough to distinguish between the spectral shapes of $\mathcal{O}_3$ and $\mathcal{O}_1$. Si (also Fig.~\ref{fig:op3_simulated_data}) has a low number of simulated events such that little information on the spectral shape is available. However, the relatively large range of recoil energies points to the nature of the underlying spectrum; for an exponentially decaying spectrum, most of the WIMP events would be expected to cluster at the experimental threshold. 

\begin{table}[t]
\begin{tabular}{c|c|c|c}
Reconstructed point & $m_\chi$ (GeV) & $\{c_1^0,c_1^1\}\times10^3$ & $\{c_3^0,c_3^1\}$ \\ \hline \hline
\multirow{2}{*}{BP$_3$ in $\mathcal{O}_1$} & $11.8$ & $\{0.26,5.4\}$ & \multirow{2}{*}{...} \\
& $(9.7,14.6)$ & $\{(0.15,0.45),(-0.9,10.4)\}$ & \\ \hline
\multirow{2}{*}{BP$_3$ in $\mathcal{O}_3$} & $8.1$ & \multirow{2}{*}{...} & $\{14.2,-8\}$ \\
& $(7.0,9.3)$ & & $\{(7.2,19.2),(-51,-27)\cup(-24,3)\}$ \\ \hline
\multirow{2}{*}{BP$_3$ in $\mathcal{O}_1$ and $\mathcal{O}_3$ }& $8.1$ & $\{0.13,0.5\}$ & $\{15,-8\}$ \\
& $(6.9,9.7)$ & $\{(0.00,0.42),(-5.3,4.2)\}$ & $\{(-21,0)\cup(4,28),(-62,-29)\cup(-27,54)\}$ \\ \hline
\end{tabular}
\caption{Best fit points with 95\% confidence regions for the 3D and 5D reconstructions of the benchmark point BP$_3$ of Table \ref{tab:sim_parameters}, based on 1D marginalized likelihoods. As noted in the header, $\mathcal{O}_1$ coupling coefficients have been enlarged by $10^3$.}
\label{tab:best_fit_points_3}
\end{table}

\begin{figure}[t]
\centering
\begin{subfigure}{\columnwidth}
\includegraphics[width = 0.875\columnwidth]{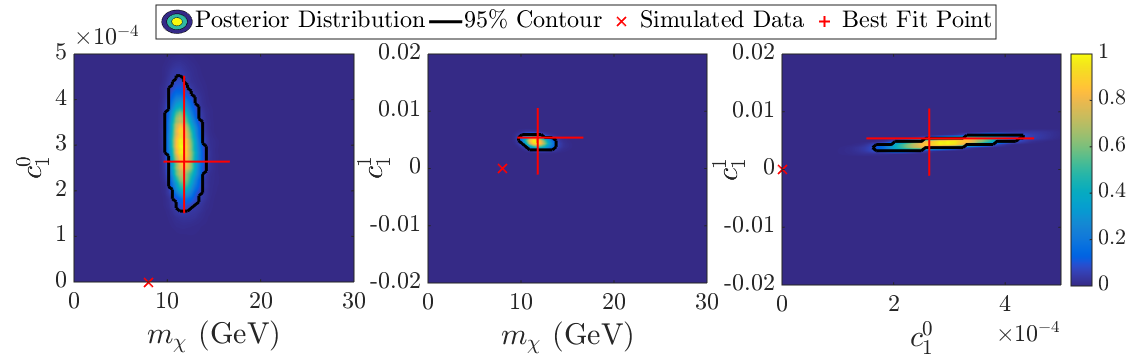}
\includegraphics[width = 0.875\columnwidth]{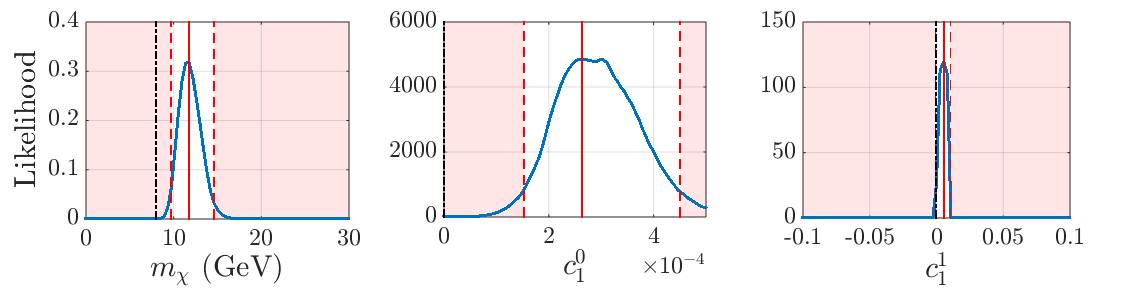}
\caption{$\mathcal{O}_1$ recovery}
\label{fig:op3_3Dlikelihoods_top}
\end{subfigure}
\begin{subfigure}{\columnwidth}
\includegraphics[width = 0.875\columnwidth]{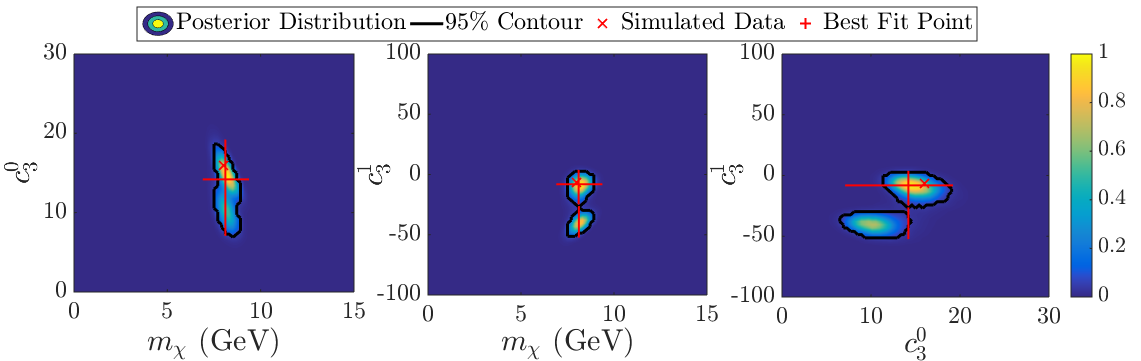}
\includegraphics[width = 0.875\columnwidth]{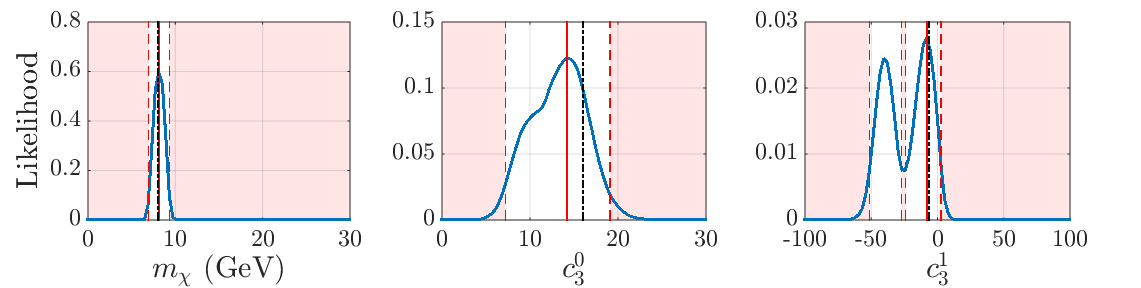}
\caption{$\mathcal{O}_3$ recovery}
\label{fig:op3_3Dlikelihoods_bottom}
\end{subfigure}
\caption{3D likelihoods of the data simulated using BP$_3$ and analyzed under the assumption of $\mathcal{O}_1$ (\ref{fig:op3_3Dlikelihoods_top}) or $\mathcal{O}_3$ (\ref{fig:op3_3Dlikelihoods_bottom}) for all three experiments combined. For each recovery operator, the top row of plots shows 2D marginalized likelihoods (obtained by marginalizing over one of the parameters) and the bottom row shows the 1D marginalized likelihoods (obtained by marginalizing over two of the three parameters). Also shown is the point representing the simulated data, marked by x in 2D and a black dashed line in 1D and the best fit point represented by the red + in 2D and red vertical line in 1D.}
\label{fig:op3_3Dlikelihoods}
\end{figure}

As in the case of $\mathcal{O}_8$ above, we use 3D EFT likelihood analyses to test steps \ref{item:3D} and \ref{item:Bayes} of the proposed analysis procedure. The 3D likelihood is first computed assuming that the WIMP-nucleon scattering proceeds via the standard SI operator, that is over the parameter space $\{m_{\chi},c_1^0,c_1^1\}$. This is then contrasted with the likelihood computed assuming the correct scattering operator, that is over the parameter space $\{m_{\chi},c_3^0,c_3^1\}$. Both likelihoods are joint, combining all three simulated experiments (Si, Ge, and LXe). Figure \ref{fig:op3_3Dlikelihoods_top} shows the 2D marginalized likelihoods (top) and the 1D marginalized likelihoods (bottom) assuming the $\mathcal{O}_1$ interaction. In both the 2D and 1D marginalized likelihoods, the simulated data point represented by $\{m_\chi, c_1^0,c_1^1\}=\{8.0\text{~GeV},0,0\}$ is not included in the 95\% confidence contours/intervals. That is, these contours do not accurately represent the underlying nature of the simulated dark matter, which is a consequence of assuming the wrong interaction operator in the analysis. The parameter values of the point of maximum likelihood with 95\% confidence intervals calculated from the 1D marginalized likelihoods are shown in Table \ref{tab:best_fit_points_3}.

If instead the analysis assumes the same operator as the simulation (in this case $\mathcal{O}_3$), the 95\% confidence contours include the simulated data point $\{m_\chi, c_3^0, c_3^1\}=\{8.0\text{~GeV}, 16, -6.4\}$ as shown in Fig.~\ref{fig:op3_3Dlikelihoods_bottom} by the 2D marginalized likelihoods (top) and 1D marginalized likelihoods (bottom). Even though two regions of high likelihood are visible in each 2D marginalized likelihood, the likelihood favors the region that contains the simulated data point. Additionally, the point of maximum likelihood agrees closely with the simulated data point. This is also shown numerically in Table \ref{tab:best_fit_points_3}. Bayesian evidence further supports the hypothesis that the operator $\mathcal{O}_3$ fits the simulated data better than $\mathcal{O}_1$: the evidence calculated for $\mathcal{O}_3$ is $2 \times10^{-18}$, about 20 times higher than for $\mathcal{O}_1$ at $1\times10^{-19}$. This indicates that the simple Bayesian evidence measure can be used to compare recoveries with different assumed operators in order to determine which operator(s) perform best in terms of explaining the observed data from multiple experiments.

Both 3D likelihoods shown in Fig.~\ref{fig:op3_3Dlikelihoods} were calculated by combining all three experiments into a single likelihood, resulting in better-defined contours than for any individual experiment. Even when the likelihood for each experiment individually is fairly flat over the entire prior range, such as for $c_3^0$ from the 3D likelihood assuming $\mathcal{O}_3$ interaction, combining experiments can create a closed contour for the coupling coefficient, as shown in Fig.~\ref{fig:op3_3Dlikelihoods_bottom}. Unfortunately, it is also possible to obtain closed contours when combining experiments for a likelihood calculated by assuming the wrong EFT operator, which stresses the importance of considering the Bayesian evidence.

The Ge and LXe experiments detected many more events than Si, with 19 and 21 respectively, compared to only three for Si. The 95\% confidence intervals calculated for the joint likelihood assuming $\mathcal{O}_3$ interaction, shown in Table \ref{tab:best_fit_points_3}, are, on average, 1.4 times tighter than for Ge only and 4.3 times tighter than for LXe only. Since there were more simulated data events in LXe than in Ge, it might be expected that the LXe contours would be the closest to the joint likelihood. However, due to the flatness of the $\mathcal{O}_3$ likelihood in LXe for the isovector coupling coefficient, $c_3^1$, the average between the three 2D marginalized likelihoods is slightly higher than for Ge alone or for Si, Ge, and LXe combined.

\begin{figure}[t]
\centering
\includegraphics[width = \columnwidth]{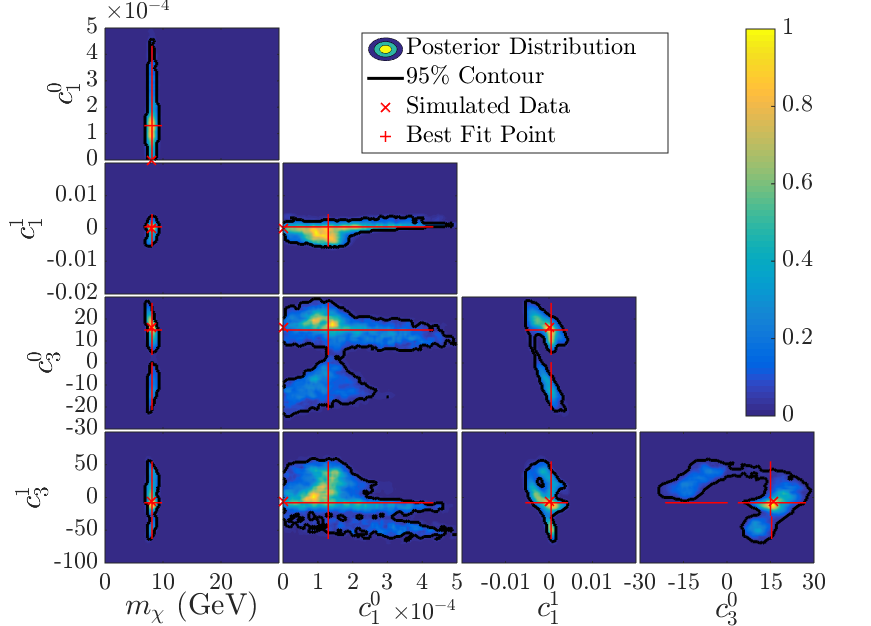}
\includegraphics[width = \columnwidth]{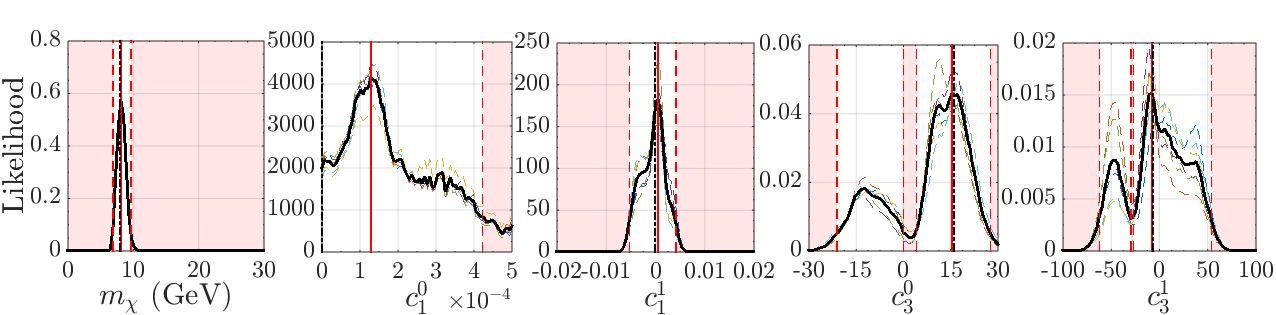}
\caption{5D likelihood of the data simulated in $\mathcal{O}_3$ and computed assuming WIMP-nucleon interaction in $\mathcal{O}_1$ and $\mathcal{O}_3$, for all three experiments combined. The 95\% confidence contours in 2D marginalized likelihoods are shown on the top and the 1D marginalized likelihoods are shown in the bottom row of plots. The multiple colors in the 1D plot represent the marginalizations of the six subspaces and the black line the averaged. Also shown is the point representing the simulated data, marked by x in 2D and a black dashed line in 1D and the best fit point represented by the red + in 2D and red vertical line in 1D. }
\label{fig:op3_5Dlikelihoods}
\end{figure}

As suggested in step \ref{item:5D} of the proposed analysis procedure, computing the 5D likelihood for both $\mathcal{O}_1$ and $\mathcal{O}_3$ should help differentiate between the two operators by allowing constraints to be set simultaneously for both operators. Since the simulation assumed only nonzero components in $\mathcal{O}_3$, the $\mathcal{O}_1$ coupling coefficient contours should include zero, which was the simulated value of those parameters. For the $\mathcal{O}_8$ simulation, the 5D likelihood including $\mathcal{O}_1$ and $\mathcal{O}_8$ ended up being overparameterized due to the similar recoil energy spectral shapes for all four of the coupling coefficients involved and due to the low number of simulated data points. The $\mathcal{O}_3$ simulation has the advantage of having more simulated WIMP events and different spectral shapes for $\mathcal{O}_1$ and $\mathcal{O}_3$. We compute the 5D likelihood over the parameter space $\{m_{\chi},c_1^0,c_1^1,c_3^0,c_3^1\}$ and show the 2D marginalized likelihoods (top) and the 1D marginalized likelihoods (bottom) in Fig.~\ref{fig:op3_5Dlikelihoods}. The 1D marginalized likelihoods were computed in the same manner as for $\mathcal{O}_8$. Unlike the $\mathcal{O}_8$ simulation, none of the parameters in the 5D analysis have a flat likelihood. Hence, the simulated data point is better recovered, and it is fully contained within all of the 2D 95\% confidence contours and 1D 95\% confidence intervals.

Table \ref{tab:best_fit_points_3} shows the value of the point of highest likelihood with the 95\% confidence intervals as calculated from the 1D marginalized likelihoods. The simulated data point is well contained within all of the 95\% confidence intervals and is closer to the point of maximum likelihood than suggested by the width of these intervals. In this instance, the 5D likelihood was able to successfully fit the simulated values of WIMP mass and coupling coefficients. The point of highest likelihood is very similar to that of the 3D $\mathcal{O}_3$ likelihood, but as expected by the increase in the number of parameters, the 95\% confidence intervals are larger for the 5D than for the 3D likelihood.

\section{Conclusion}\label{sec:conclusions}

With the systematic analysis procedure suggested here, higher-dimensional analysis of dark matter data using the model-independent EFT framework is possible. \textsc{MultiNest} \cite{Feroz:2008xx,Feroz:2007kg,Feroz:2013hea} is an effective Bayesian inference tool that can be used to efficiently scan high-dimensional likelihoods. The number of operators or dimensions in a likelihood can be limited by using the Bayesian evidence for single-operator 3D likelihoods to determine which operators best fit the observed events. Higher-dimensional likelihoods can be marginalized down to 2D and 1D likelihoods in order to ease visualization and set constraints on the WIMP mass and coupling coefficients.

Assumptions about the operators could lead to tension between experiments. For example, the tension between the isoscalar operator 1 (spin-independent) analyses published by CDMS II Si~\cite{Agnese:2013rvf}, CDMS II Ge~\cite{Agnese:2015ywx}, and LUX~\cite{Akerib:2013tjd} could be relieved by including other coupling coefficients in the analysis, such as the isovector operator 1 component, while setting new limits on dark matter interactions. Combining the three experiments into a single joint likelihood leads to stronger limits than what is possible from a single target or experiment alone.

Using simulated data (assuming $\mathcal{O}_3$ or $\mathcal{O}_8$ interaction) to test the proposed analysis procedure showed that the simulated data point can be reconstructed in both 3D and 5D likelihood analyses. Comparisons of the Bayesian evidence for 3D (WIMP mass and a single EFT operator) likelihoods can identify which operator(s) fit the data well. However, it is critical to include more than one target in the analysis, in order to differentiate between operators of similar recoil energy spectra and to create better defined confidence contours, especially when dealing with a low number of detected events per experiment.

When operators, such as $\mathcal{O}_1$ and $\mathcal{O}_3$, have different recoil energy spectral shapes due to different momentum dependencies, they can be more easily distinguished from each other by the proposed analysis procedure. When using spectral shape in this way, it is extremely important to have a low enough experimental energy threshold in order to be able to measure the spectral differences. For a recoil energy spectrum similar to that of $\mathcal{O}_3$, if the experimental energy threshold is above or near the peak of the spectrum (as it was for LXe in our $\mathcal{O}_3$ simulation) then the spectral shape can appear to follow the standard exponential decay of $\mathcal{O}_1$. Very low-threshold dark matter experiments, such as the previously published CDMSlite \cite{Agnese:2013jaa,Agnese:2015nto} and the proposed SuperCDMS SNOLAB high-voltage experiments \cite{Agnese:2016cpb} will be particularly useful to convincingly perform spectral shape discrimination in the EFT framework.

\begin{acknowledgments}
The authors gratefully acknowledge Blas Cabrera, Steve Yellin, Wolfgang Rau, Ray Bunker, Rob Calkins, Mark Pepin, Kristi Schneck, and the rest of the SuperCDMS collaboration for their discussions of this work. This work is supported in part by the United States Department of Energy (DE-SC0012294), STFC (U.K.), MultiDark (CSD2009-0006), and the Severo Ochoa Program (SEV-2012-0249).
\end{acknowledgments}

\bibliographystyle{apsrev4-1}
\bibliography{EFT_bibliography}

\end{document}